\documentclass[twocolumn,aps,prl,superscriptaddress]{revtex4-1}
\usepackage{graphicx,subfigure}
\usepackage{amsmath,amssymb,bm}
\usepackage{mathtools}
\usepackage{multirow}
\usepackage{makecell}
\usepackage{textcomp}
\usepackage{float}
\usepackage{xcolor, soul}

\usepackage{color}
\usepackage[normalem]{ulem}
\usepackage{dsfont}
\usepackage{mathrsfs}
\usepackage{braket}
\usepackage{transparent}
\usepackage{xcolor}
\usepackage{hhline}
\usepackage{graphicx}
\usepackage{pdfpages}
\usepackage{bibunits}
\usepackage{filecontents,hyperref}

\makeatletter
\AtBeginDocument{\let\LS@rot\@undefined}
\makeatother

\usepackage{graphicx}

\graphicspath{ {figures/} }

\makeatletter
\let\saved@includegraphics\includegraphics
\AtBeginDocument{\let\includegraphics\saved@includegraphics}
\makeatother

\newcommand{\figOne}{
 \begin{figure}[t]
    \centering
    \includegraphics[width=\columnwidth]{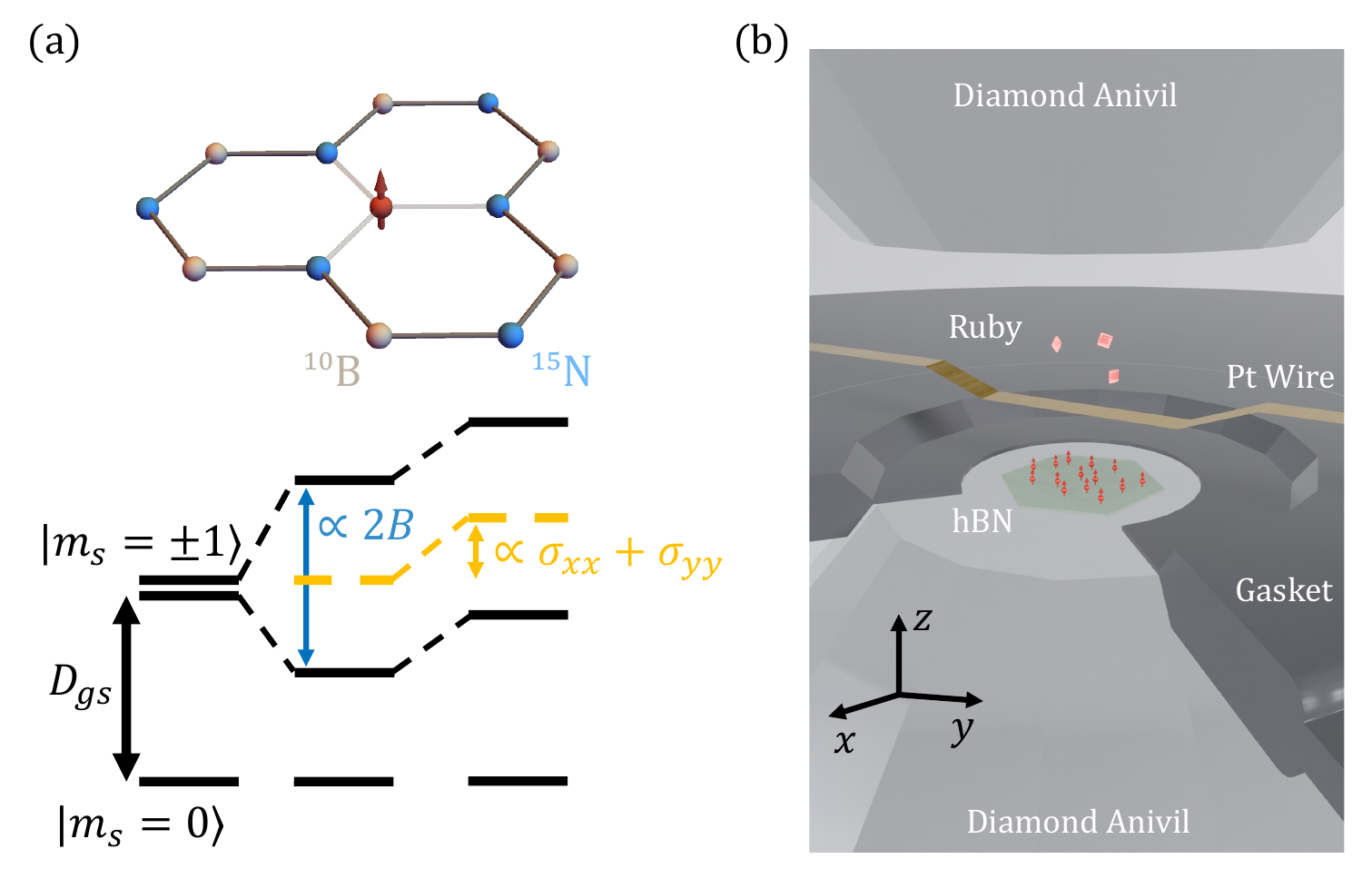}
    \caption{{\bf Experimental platform for high-pressure quantum sensing using \vbm defects in hBN.} (a) Top: Schematic of the hBN layer with \vbm defects used for stress and magnetism sensing. Bottom: Energy-level diagram of the \vbm spin triplet ground state, showing Zeeman splitting induced by a magnetic field and energy shifts caused by in-plane stress components. The dual sensing capabilities enable simultaneous characterization of stress and magnetic fields. (c) Cross-sectional schematic of the DAC, illustrating the positioning of the hBN flake containing \vbm defects on the culet, the platinum wire for microwave delivery, and the ruby microsphere for pressure calibration.
    }
    \label{fig:fig1}
\end{figure}
}

\newcommand{\figTwo}{
 \begin{figure}[t]
    \centering
    \includegraphics[width=\columnwidth]{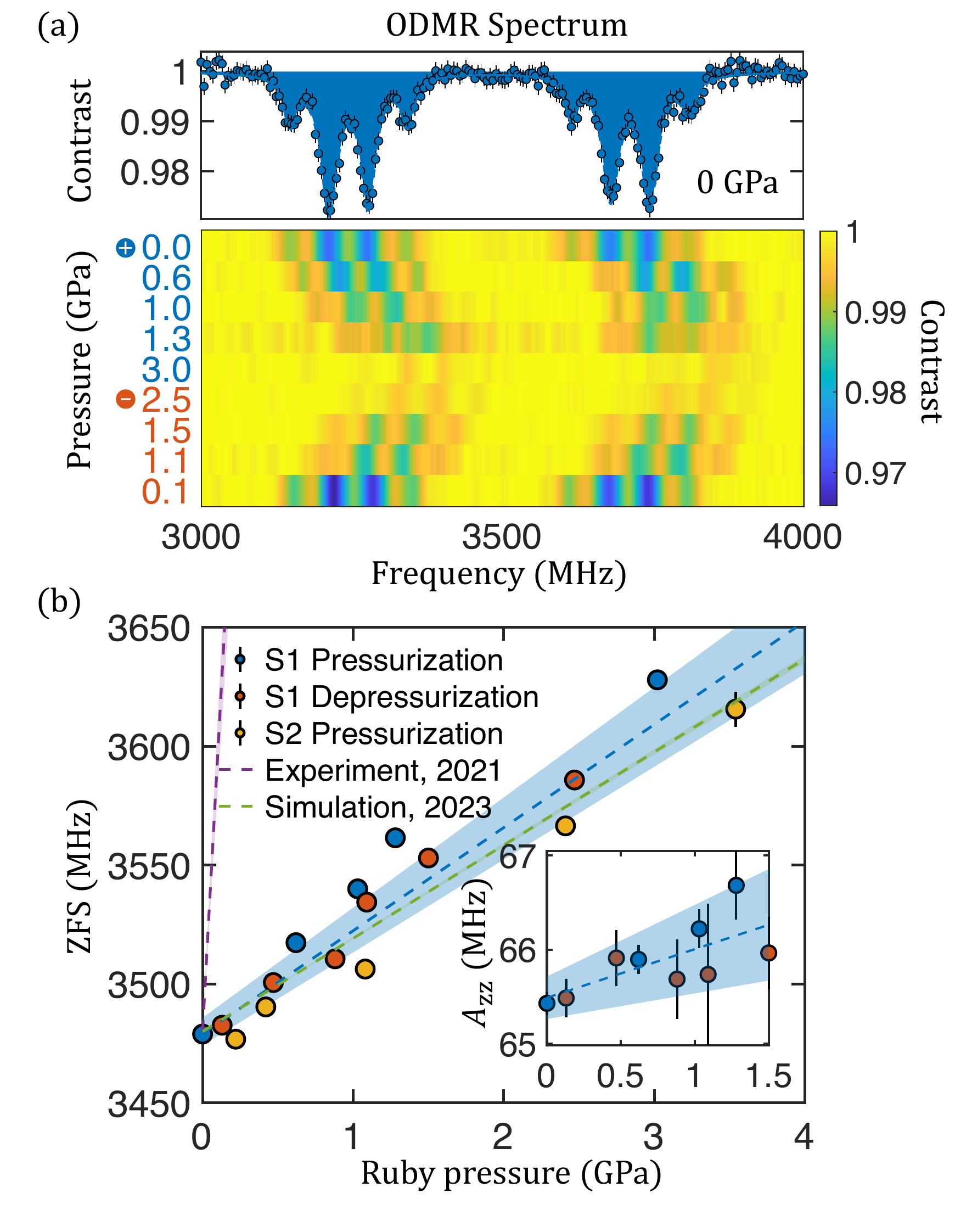}
    \caption{{\bf ODMR spectra of \vbm defects under varying pressures.} (a) Top: A representative ODMR spectrum at $0~$GPa, with the shaded area indicating a fit to the spectrum using a sum of 8 Lorentzian functions. Bottom: A waterfall plot demonstrating a global shift in resonance frequencies with increasing (blue) and decreasing (red) pressure, reflecting the \vbm sensitivity to lattice strain. (b) Extracted center frequencies of the \vbm spin transitions as a function of pressure, independently calibrated using ruby R2 fluorescence shifts. Data points for two iterations are shown: pressurization for Sample 1 (blue circles) and Sample 2 (yellow circles), and depressurization for Sample 1 (red circles). A linear fit (blue dashed line) of the two rounds of measurements yields a pressure susceptibility of $(2\pi)\times(43 \pm 7)~$MHz/GPa. This susceptibility is consistent with a theoretical prediction (green) but significantly deviates from a previous experiments (purple). Shaded regions denote one standard deviation of fits. Inset: Hyperfine splitting increases with pressure, yielding a susceptibility of $(2\pi)\times(0.5\pm0.2)~$MHz/GPa.
    }
    \label{fig:fig2}
\end{figure}
}

\newcommand{\figThree}{
 \begin{figure*}[t]
    \centering
    \includegraphics[width=0.8\textwidth]{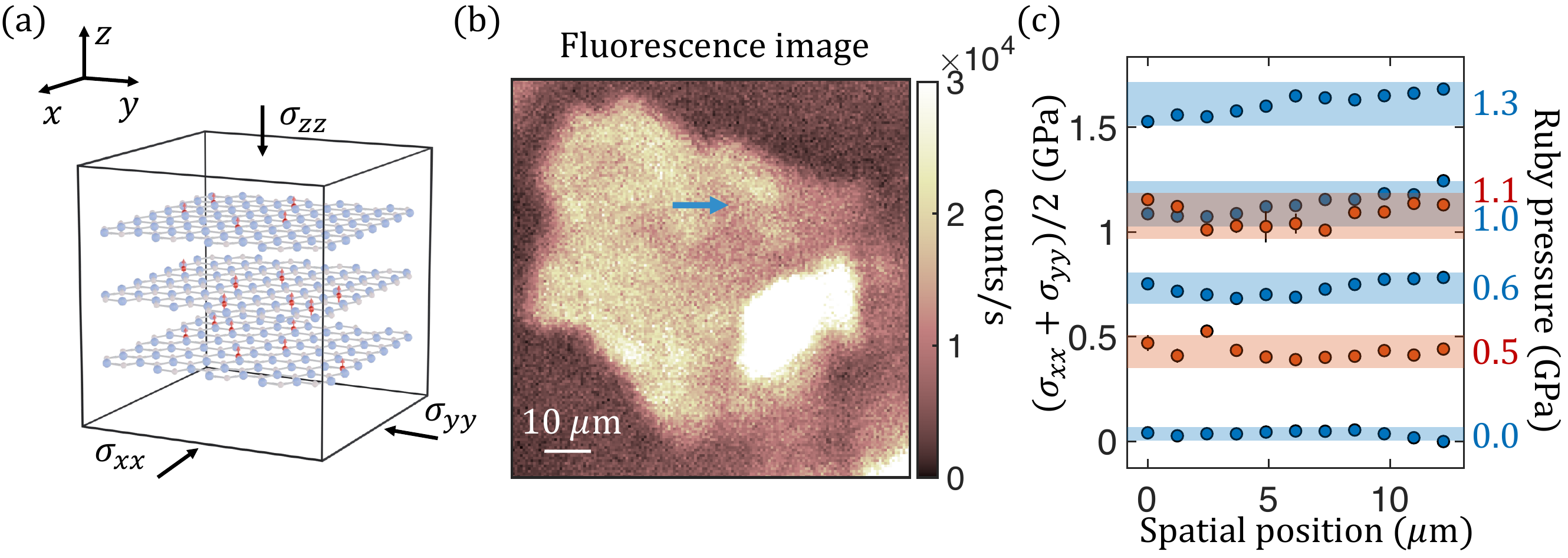}
    \caption{{\bf Spatial profiles of stress gradients within the high-pressure chamber.} (a) A schematic illustrating the layered atomic structure of hBN. This structure enables \vbm to function as an in-plane hydrostatic pressure sensor by coupling strongly to lateral stress components while remaining insensitive to out-of-plane stress. (b) A fluorescence scanning image of the hBN flake in the DAC under pressure, with a linecut direction marked by the blue arrow for stress profiling. (c) Measured lateral stress profile along the linecut shown in (b) for varying pressures, including pressurization (blue) and depressurization (red). At low pressure, the stress distribution is uniform, but as pressure increases beyond 1 GPa, significant stress gradients develop due to the non-hydrostaticity of the solid NaCl pressure medium and slight misalignments of the diamond anvils.
    }
    \label{fig:fig3}
\end{figure*}
}

\newcommand{\figFour}{
 \begin{figure}[t]
    \centering
    \includegraphics[width=\columnwidth]{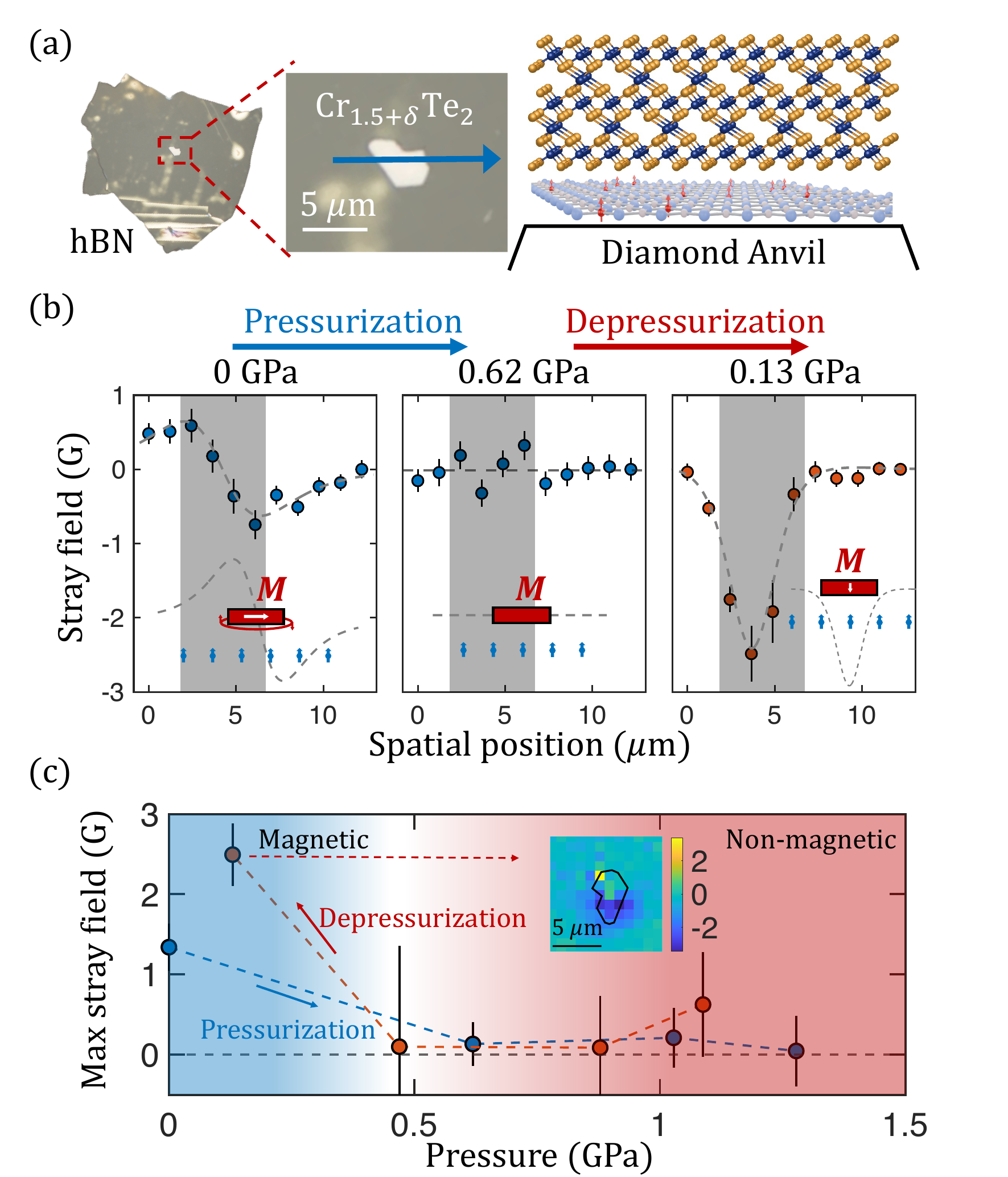}
    \caption{{\bf Investigation of pressure-induced magnetic transitions in self-intercalated Cr$_{1+ \delta}$Te$_2$ using \vbm defects.} (a) A schematic of the DAC assembly, illustrating the heterostructure including Cr$_{1+ \delta}$Te$_2$ sample on top of a hBN flake, and a cross-sectional view of the crystal structure. (b) Spatial profile of stray magnetic fields across Cr$_{1+ \delta}$Te$_2$ at $0~$GPa, $0.62~$GPa, and $0.13~$GPa. Insets show theoretical expectations of the stray field detected by a \vbm located adjacent to a magnetic sample. (c) Evolution of the maximum stray magnetic field difference as a function of pressure, with data showing pressurization (blue) and depressurization (red) cycles. The maximum field is determined by calculating the stray field difference between the two edges of the sample during the pressurization process. For depressurization, it is calculated based on the difference between the stray field at the center of the sample and that at a distant point. Inset: A two-dimensional scan of the stray field around Cr$_{1+ \delta}$Te$_2$ sample at $0.13~$GPa. 
    }
    \label{fig:fig4}
\end{figure}
}
\newcommand{\figAFM}{
 \begin{figure}[h]
    \centering
    \includegraphics[width=0.5\columnwidth]{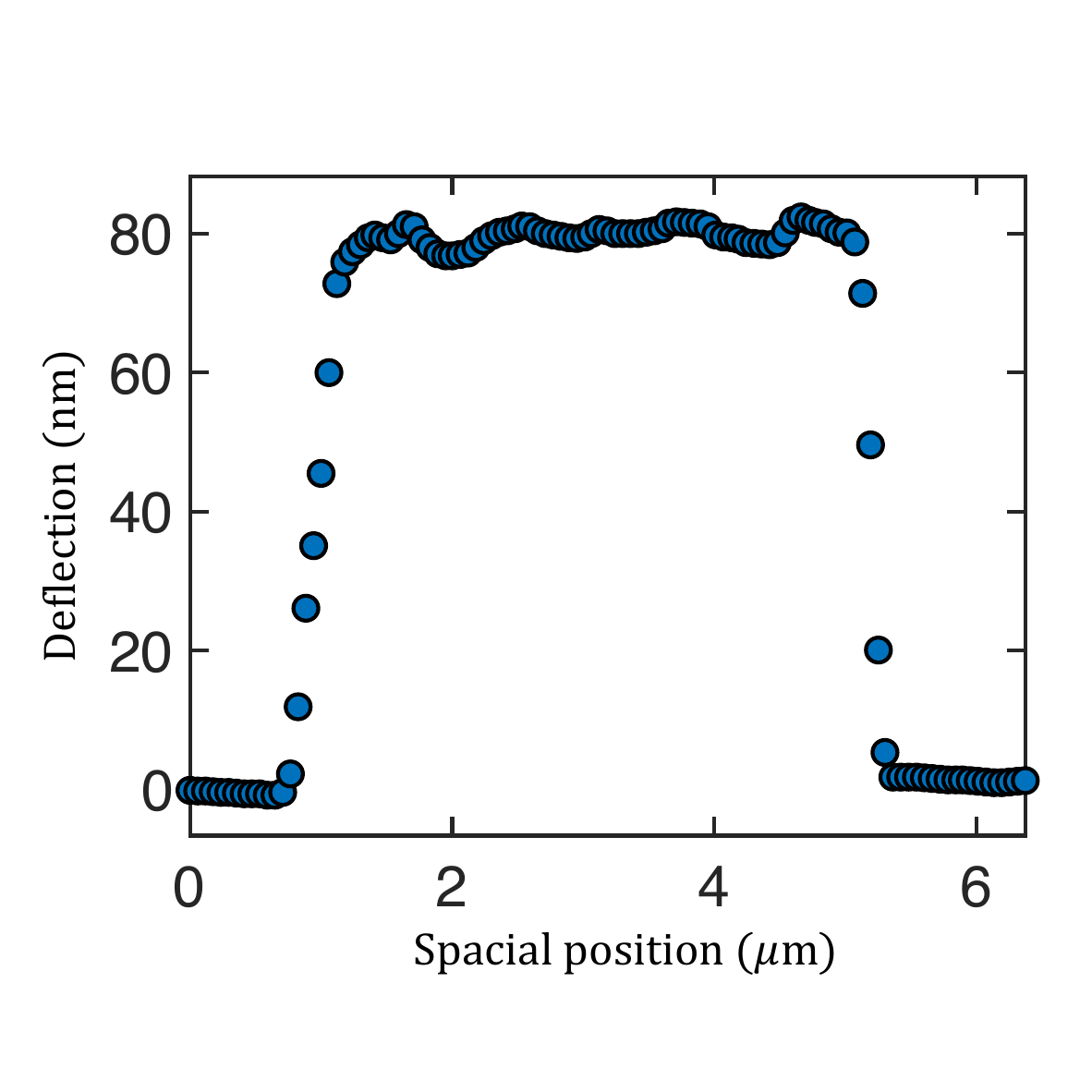}
    \caption{Linecut deflection measurement across the Cr$_{1+ \delta}$Te$_2$ sample obtained using atomic force microscopy (AFM). The calibrated thickness of flakes synthesized in the same batch ranges from 50–100~nm.
    }
    \label{fig:AFM}
\end{figure}
}
\newcommand{\figTEM}{
 \begin{figure}[h]
    \centering
    \includegraphics[width=\columnwidth]{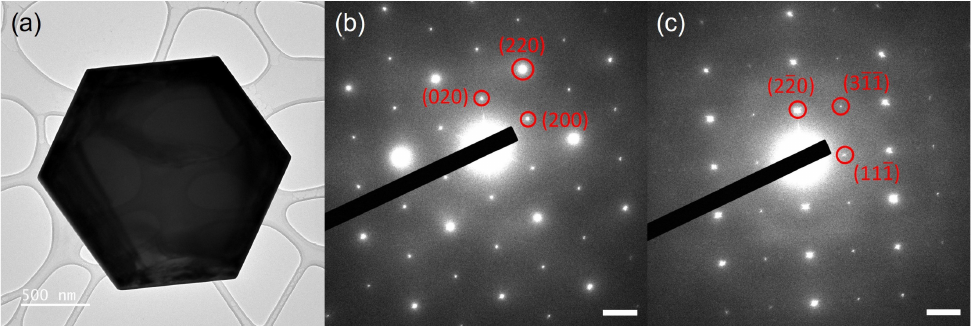}
    \caption{(a) TEM image of a chromium telluride nanoplate. (b) and (c) corresponding SAED patterns along zone axes [001] and [112]. The scale bars in (b) and (c) are 2~nm$^{-1}$.
    }
    \label{fig:TEM}
\end{figure}
}

\newcommand{\figDFT}{
 \begin{figure}[h]
    \centering
    \includegraphics[width=\columnwidth]{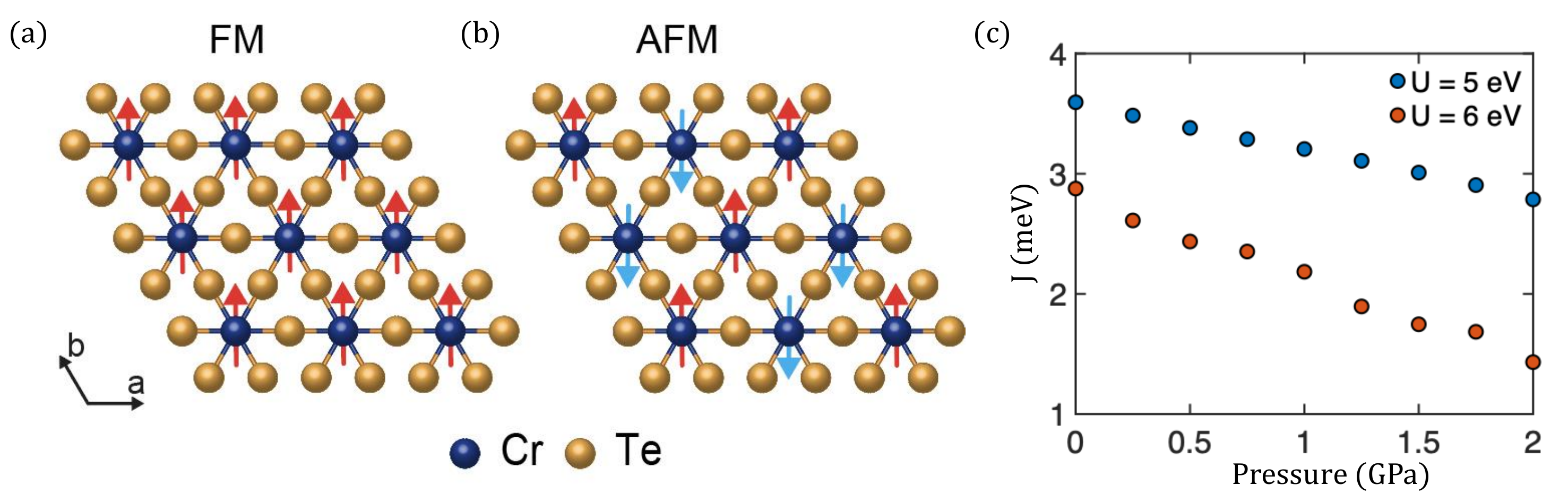}
    \caption{Schematic configuration of in-plane (a) ferromagnetic and (b) antiferromagnetic configurations. (c) Calculated exchange constant $J$ as a function of pressure for two different $U$ values. The blue and red spheres represent $U = 5$~eV and $U = 6~$eV, respectively.
    }
    \label{fig:DFT}
\end{figure}
}

\makeatletter
\newcommand*{\centerfloat}{
  \parindent \z@
  \leftskip \z@ \@plus 1fil \@minus \textwidth
  \rightskip\leftskip
  \parfillskip \z@skip}
\makeatother

\usepackage{lineno} 
\begin{document}

\newcommand{\vbm}[0]{$\mathrm{V}_{\mathrm{B}}^-$ }
\newcommand{\vbmns}[0]{$\mathrm{V}_{\mathrm{B}}^-$}

\newcommand{\rg}[1]{{\color{orange} RG: #1}}
\newcommand{\cz}[1]{{\color{blue} CZ: #1}}
\newcommand{\ghe}[1]{{\color{red} GHe: #1}}

\title{Probing Stress and Magnetism at High Pressures with Two-Dimensional Quantum Sensors}

\author{Guanghui~He,$^{1,*}$
Ruotian~Gong,$^{1,*}$
Zhipan~Wang,$^{2,*}$
Zhongyuan~Liu,$^{1}$
Jeonghoon~Hong,$^{3}$
Tongxie~Zhang,$^{3}$
Ariana~L.~Riofrio,$^{4}$
Zackary~Rehfuss,$^{1}$
Mingfeng~Chen,$^{1}$
Changyu~Yao,$^{1}$
Thomas~Poirier,$^{5}$
Bingtian~Ye,$^{6}$
Xi~Wang,$^{1}$
Sheng~Ran,$^{1}$
James~H.~Edgar,$^{5}$
Shixiong~Zhang,$^{3}$
Norman~Y.~Yao,$^{2}$
Chong~Zu,$^{1,7,8,\dag}$
\\
\normalsize{$^{1}$Department of Physics, Washington University, St. Louis, MO 63130, USA}\\
\normalsize{$^{2}$Department of Physics, Harvard University, Cambridge, MA 02138, USA}\\
\normalsize{$^{3}$Department of Physics, Indiana University Bloomington, Bloomington, IN 47405, USA}\\
\normalsize{$^{4}$Department of Physics, Santa Clara University, Santa Clara, CA 95053, USA}\\
\normalsize{$^{5}$Tim Taylor Department of Chemical Engineering, Kansas State University, Manhattan, KS 66506, USA}\\
\normalsize{$^{6}$Department of Physics, Massachusetts Institute of Technology, Cambridge, MA 02139, USA}\\
\normalsize{$^{7}$Center for Quantum Leaps, Washington University, St. Louis, MO 63130, USA}\\
\normalsize{$^{8}$Institute of Materials Science and Engineering, Washington University, St. Louis, MO 63130, USA}\\
\normalsize{$^*$These authors contributed equally to this work.}\\
\normalsize{$^\dag$To whom correspondence should be addressed: zu@wustl.edu}\\
}

\begin{abstract}

Pressure serves as a fundamental tuning parameter capable of drastically modifying all properties of matter.
The advent of diamond anvil cells (DACs) has enabled a compact and tabletop platform for generating extreme pressure conditions in laboratory settings.
However, the limited spatial dimensions and ultrahigh pressures within these environments present significant challenges for conventional spectroscopy techniques.
In this work, we integrate optical spin defects within a thin layer of two-dimensional (2D) materials directly into the high-pressure chamber, enabling an in situ quantum sensing platform for mapping local stress and magnetic environments up to 4~GPa. 
Compared to nitrogen-vacancy (NV) centers embedded in diamond anvils, our 2D sensors exhibit around three times stronger response to local stress and provide nanoscale proximity to the target sample in heterogeneous devices.
We showcase the versatility of our approach by imaging both stress gradients within the high-pressure chamber and a pressure-driven magnetic phase transition in a room-temperature self-intercalated van der Waals ferromagnet, Cr$_{1+ \delta}$Te$_2$. 
Our work demonstrates an integrated quantum sensing device for high-pressure experiments, offering potential applications in probing pressure-induced phenomena such as superconductivity, magnetism, and mechanical deformation.
\end{abstract}
\date{\today}
\maketitle

\emph{Introduction} --- A key requirement for developing modern integrated quantum-sensing devices is the seamless incorporation of sensors into existing toolsets. 
In high-pressure science, the recently introduced nitrogen-vacancy (NV) centers in diamond anvil cells (DAC) serve as a compelling demonstration, enabling in situ characterization of stress and magnetism under extreme pressures for various applications in materials science and geology~\cite{doherty2014electronic,hsieh2019imaging,bhattacharyya2024imaging,lesik2019magnetic,yip2019measuring,wang2024imaging,steele2017optically,shang2019magnetic,hamlin2019extreme,shelton2024magnetometry,ho2020probing,dai2022optically, rovny2024nanoscale,hilberer2023enabling,wen2024probing, wang2024imaging}.
However, limitations remain across different implementations: NV layers implanted beneath the culet of the diamond anvil are located outside the sample chamber, restricting proximity to the target. In contrast, nanodiamonds placed inside the chamber offer closer access to the sample. Yet, in this geometry, the random positions and crystalline orientations of individual nanodiamond particles hinder their suitability for widefield imaging.

To address these challenges, we propose the negatively charged boron-vacancy (\vbmns) centers in layered hexagonal boron nitride (hBN) as an alternative high-pressure sensing platform inside DACs.
\vbm centers together with a few other recently discovered spin defects in van der Waals materials 
present a brand-new angle to construct quantum sensing devices~\cite{vaidya2023quantum, azzam2021prospects, ren2019review, su2022tuning,scholten2024multi,healey2023quantum, aharonovich2022quantum, gottscholl2020initialization, gottscholl2021room, gong2023coherent, naclerio2023review, durand2023optically,stern2022room,li2022carbon, gao2023nanotube,kumar2022magnetic,das2024quantum,lyu2022strain,zabelotsky2023creation,gao2021high,huang2021two, zhou2024sensing}.
In principle, the atomic-thin structure of the host materials can enable the 2D sensors to be placed sub-nanometer away from the target samples, facilitating the imaging of inter-facial phenomena with unprecedented sensitivity and resolution. 
Moreover, 2D sensors can be easily integrated with other 2D devices through three-dimensional heterogeneous integration, allowing the large-scale manufacture of next-generation microelectronics.

In this work, we present three main results.
First, by transferring a thin  ($\sim 100~$nm) film of hBN containing \vbm centers directly onto the culet of the diamond anvil (Fig.~\ref{fig:fig1}), we systematically characterize the electronic spin properties of \vbm with pressure up to 4~GPa.
The pressure-induced spin energy shift of \vbm is measured to be $(2\pi)\times (43\pm7)$~MHz/GPa, in good agreement with first-principles calculation~\cite{udvarhelyi2023planar}, yet significantly different from a previous experiment where weights are placed directly onto the hBN~\cite{gottscholl2021spin}.
Notably, the pressure response of \vbm is around three times larger than that of NV centers in diamond, highlighting its potential as an \emph{in situ} pressure and stress sensor.
Second, we use \vbm sensors to directly map out the stress distribution and gradient inside the high-pressure chamber.
Using sodium chloride (NaCl) as the pressure-transmitting medium, we find that the stress environment becomes less uniform at around 2~GPa, consistent with previous studies~\cite{celeste2019hydrostaticity,ysj2009hydrostaticity,tateiwa2010appropriate}.
Finally, we demonstrate the magnetic field imaging capability of \vbm in a heterogeneous device by investigating the of pressure-induced magnetism in a room-temperature self-intercalated van der Waals ferromagnet, Cr$_{1+\delta}$Te$_2$.
A transition from ferro-magnetic to non-magnetic behavior occurs around $0.5$~GPa, which can be explained by a decrease in the exchange interaction.

\emph{Experimental platform} ---
We choose to work with isotopically purified h$^{10}\text{B}^{15}\text{N}$ to minimize the nuclear spin noise and achieve the optimal \vbm spin properties~\cite{gong2024isotope, clua2023isotopic, janzen2024boron,sasaki2023nitrogen}.
An ensemble of \vbm is created via neutron irradiation process (see Methods).
A \vbm center harbors an electronic spin triplet ground state $|m_s = 0,~\pm1\rangle$, which can be optically initialized and read out at room temperature.
In the absence of any external perturbations, the 
$|m_s=\pm1\rangle$ states are degenerate and separated from $|m_s = 0\rangle$ by a zero-field splitting (ZFS), $D_\mathrm{gs} = (2\pi)\times3.48~$GHz (Fig.~1a). 
An applied external magnetic field lifts the degeneracy between $|m_s=\pm1\rangle$ through Zeeman splitting, while hydrostatic pressure  induces a global energy shift of the ZFS, allowing \vbm to simultaneously probe the local magnetic and pressure environment.

\figOne

In our experiment, we use a miniature Pasternak DAC~\cite{sterer1990multipurpose} with two opposing anvils ($400~\mu$m culet diameter) compressing a stainless steel gasket (Fig.~\ref{fig:fig1}b).
To integrate \vbm sensors into the DAC, we directly transfer a thin h$^{10}\text{B}^{15}\text{N}$ flake with thickness $\sim 100~$nm onto the culet of a anvil (Fig.~\ref{fig:fig1}b). 
A coherent microwave field is delivered using a $50~\mu$m width platinum foil across the culet to control the \vbm spin states.
The sample chamber is packed with NaCl pressure-transmitting medium to provide a hydrostatic environment.
A ruby microsphere is loaded on the opposite side of the foil as a standard pressure calibrant. 
\figTwo

\emph{\vbm under pressure} --- We begin by probing the \vbm spin states under pressure via optically detected magnetic resonance (ODMR) spectroscopy: by sweeping the frequency of the applied microwave field while detecting the fluorescence signal of \vbm, we observe a fluorescence drop when the microwave is in resonance with one of the the \vbm spin transitions. 
We apply a small external magnetic field $B_\textbf{ext} = 84~$G along the loading axis of the DAC (i.e., the out-of-plane direction of hBN) to lift the degeneracy between $|m_s = \pm1\rangle$ sublevels.
The results are shown in Fig.~\ref{fig:fig2}a.
At 0~GPa, two groups of electronic spin transitions are observed symmetrically around \( D_\mathrm{gs} = (2\pi) \times 3.48~\mathrm{GHz} \), corresponding to \( |m_s = 0\rangle \longleftrightarrow |m_s = \pm 1\rangle \).
Each electronic spin transition further exhibits four hyperfine resonances with a coupling strength of \( A_\mathrm{zz} = (2\pi) \times 65.5~\mathrm{MHz} \), attributed to the hyperfine interaction between the \vbm spin and three nearby \( ^{15}\mathrm{N} \) nuclei~\cite{gong2023coherent,gong2024isotope,plo2024isotope}.

As pressure increases~(Fig.~\ref{fig:fig2}a), the ODMR resonances exhibit a systematic shift to higher frequencies. 
This upward shift can be intuitively understood as a consequence of pressure-induced compression of the \vbm electronic spin wavefunction, leading to a enhanced spin-spin interactions.
However, the ODMR contrast decreases with increasing pressure, eventually vanishing at approximately 4~GPa.
Upon depressurization, the resonant peaks return to their original positions, and the contrast is fully restored to its initial levels.
The reduced contrast may be due to the quasi-hydrostatic nature of the NaCl pressure medium at 4~GPa, which could influence the inter-system crossing rates of \vbmns.
Employing a more hydrostatic pressure medium, such as Argon or Neon gas, may help to substantially extend the pressure limit of \vbm sensors~\cite{klotz2009hydrostatic,angel2007effective,takemura2021hydrostaticity}.

We fit the measured ODMR curves to a sum of eight Lorentzian functions to extract the ZFS at different pressures calibrated by ruby fluorescence (Fig.~\ref{fig:fig2}b).
The results includes two samples and three experimental iterations: pressurization for sample S1 (blue) and S2 (yellow), and depressurization for S1 (red). 
A linear fit across all datasets yields a pressure susceptibility of $h = (2\pi)\times(43\pm 7)~$MHz/GPa, around three times larger than that of NV centers in diamond~\cite{doherty2014electronic,barson2017nanomechanical,steele2017optically}.
For hyperfine coupling strength $A_\mathrm{zz}$, we also observe a pressure dependence of $(2\pi)\times(0.5\pm 0.2)~$MHz/GPa (Fig.~\ref{fig:fig2}b Inset).

\emph{In situ mapping of stress distribution} --- To quantitatively interpret the large pressure dependence of \vbm ZFS, let us first analyze the stress field interaction in \vbmns. 
Since each individual \vbm resides within a single layer of the hBN crystal, out-of-plane ($\hat{z}$) distortions are expected to have only a minimal effect on the in-plane lattice structure that hosts the defect's spin density~\cite{udvarhelyi2023planar}. As a result, the spin state of \vbm couples strongly to in-plane stress components $\sigma_{xx}$ and $\sigma_{yy}$, while remaining insensitive to out-of plane stress $\sigma_{zz}$ (Fig.~\ref{fig:fig3}a). The ground state Hamiltonian of the \vbm electronic degree of freedom can be then expressed as:
\begin{equation} \label{eq1}
\begin{split}
\mathcal{H}_\mathrm{gs} = D_\mathrm{gs}S_z^2 + \gamma B_\mathrm{ext}S_z + h\frac{\sigma_{xx}+\sigma_{yy}}{2}S_z^2 \\
+h'\frac{\sigma_{xx}-\sigma_{yy}}{2}(S_x^2-S_y^2),
\end{split}
\end{equation}
where $D_\mathrm{gs}=(2\pi)\times3.48~$GHz is the ZFS, $\gamma = (2\pi)\times2.8~$MHz/G is the spin gyromagnetic ratio, $B_\mathrm{ext}$ is the external magnetic field, and $S$ is the electronic spin-1 operator. $h$, $h'$, $\sigma_{xx}$ and $\sigma_{yy}$ describe the stress coupling coefficients and the lateral stress components along $\hat{x}$ and $\hat{y}$, respectively.

\figThree

Due to the presence of a pressure-transmitting medium, the stress environment within the sample chamber approximates the hydrostatic condition, i.e. $\sigma_{xx} \approx \sigma_{yy}\approx\sigma_{zz}$. In this case, the last term in Eqn.~\ref{eq1} vanishes and one only needs to consider the term, $h\frac{\sigma_{xx}+\sigma_{yy}}{2}S_z^2$, which leads to an overall energy shift of the ZFS.
From the measured pressure susceptibility, we obtain $h = (2\pi)\times (43\pm7)~$MHz/GPa, in good agreement with the previous  \emph{ab initio} calculation of $h_\mathrm{th} = (2\pi)\times39.2$~MHz/GPa~\cite{udvarhelyi2023planar}.
Notably, this value is around 25 times smaller than the result obtained from an earlier experiment, which attributed the large temperature-dependent ZFS shift to effective stress from thermal distortion of the hBN lattice~\cite{gottscholl2021spin}. However, our recent work demonstrates that the temperature-dependent ZFS change of \vbm arises from local spin-phonon interactions, rather than lattice distortion~\cite{liu2024temperature}. Consequently, the results here represent the first direct measurement of the stress response of \vbm centers. 

Having characterized the stress susceptibility of \vbm centers, we now harness \vbm sensors to perform \emph{in situ} imaging of the stress distribution within the high-pressure chamber (Fig.~\ref{fig:fig3}b). 
Fig.~\ref{fig:fig3}c shows the measured lateral stress profile across $\sim15~\mu$m region at various pressures. At $0~$GPa, the stress is uniform, reflecting the minimal applied pressure after sealing the sample chamber. As the pressure increases, a spatial gradient emerges in the stress profile and becomes markedly pronounced above $1~$GPa.
This spatial stress gradient may result from a slight misalignment of the opposing diamond anvils or a decrease in the hydrostaticity of the NaCl medium at higher pressures~\cite{ysj2009hydrostaticity}.

A few remarks are in order.
First, the stress sensitivity of \vbm sensors is estimated to be $\eta_\mathrm{p} \approx 0.2~$GPa~Hz$^{-\frac{1}{2}}$ (see Supplementary Materials). 
Although the present \vbm sensitivity is still a few times lower than that of state-of-the-art NV centers, \vbm centers offer a distinct advantage: they are directly integrated into the sample chamber. In contrast, NV centers are embedded within the diamond anvil, so one must rely on the assumption of stress continuity to infer the stress environment inside the sample chamber~\cite{hsieh2019imaging}.
Second, the primary factor limiting the sensitivity of \vbm sensors is their relatively weak fluorescence intensity. Efforts to enhance the fluorescence yield, such as by applying symmetry-breaking strain~\cite{yang2022spin, curie2022correlative, lee2024intrinsic, clua2024spin} or coupling \vbm centers to nanocavities~\cite{qian2022unveiling, fröch2021coupling, nonahal2022coupling}, could substantially improve their sensitivity.

\figFour

\emph{Imaging pressure-driven magnetism in heterogeneous device} --- To showcase the versatility of our 2D quantum sensors, we next investigate the pressure-driven magnetic phase transition in a heterogeneous device consisting of a hBN sensing layer and a thin flake of self-intercalated van der Waals magnet, Cr$_{1+\delta}$Te$_2$. Similar to the stress case, we estimate a magnetic sensitivity of \vbm to be around $\eta_\mathrm{B} \approx 3.1~$G~Hz$^{-\frac{1}{2}}$ at pressure around 0.6~GPa (see Supplementary Materials).

The recent development of van der Waals magnets has opened new avenues advanced spintronics and quantum technologies~\cite{gong2019two, gibertini2019magnetic, jiang2021recent,wang2022magnetic,burch2018magnetism}. Among a wide range of 2D magnets, transition metal telluride compounds exhibit robust ferromagnetism in ambient conditions, providing a reliable platform for practical device integration~\cite{conner2024enhanced,huang2021two,gong2017discovery,deng2018gate}.
Here we synthesize Cr$_{1+\delta}$Te$_2$ nanoflakes via a chemical vapor deposition process, where the Cr atoms are self-intercalated into the van der Waals gap between 1T-CrTe$_2$ layers~\cite{coughlin2020near,coughlin2021van}.
Transmission electron microscopy (TEM) studies indicate that the primary phase corresponds to $\delta \approx 0.5$ (see Methods).
The thickness of the nanoflakes ranges from $50-100~$nm, with Curie temperature $T_c>330~$K.
The heterostructure assembly begins with transferring a Cr$_{1+\delta}$Te$_2$ nanoflake (lateral dimension $\sim 5~\mu$m) on top of the hBN sensing layer, followed by placing the device onto the culet of a diamond anvil (Fig.~\ref{fig:fig4}a).
To minimize the oxidation of Cr$_{1+\delta}$Te$_2$, all assembly steps of the DAC are carried out in a nitrogen-filled glove box.

We directly characterize the spatial dependence of the stray magnetic fields across the Cr$_{1+\delta}$Te$_2$ nanoflake at various pressures.
All the magnetism measurement are carried out at room temperature, with an external magnetic field $B_\mathrm{ext} \approx 84~$G aligned along the out-of-plane ($\hat{z}$) axis of hBN.
Before pressurization ($P\sim0$~GPa), the stray field exhibits a characteristic spatial dependence as the measurement location scans across the sample (Fig.~\ref{fig:fig4}b).
The field initially increases, peaking at one edge, and then reverses polarity upon crossing the sample, eventually reaching a pronounced negative value at the opposite edge.
This phenomena is consistent with the expectation that the initial magnetization is oriented in-plane.
As the pressure increases above $0.5~$GPa, the stray field vanishes, indicating a transition to a non-magnetic state.

Notably, during the depressurization process, a significant stray magnetic field re-emerges when the pressure drops below $0.5~$GPa.
However, the spatial distribution of this restored stray field differs markedly from the initial configuration.
In particular, the measured magnetic field exhibits a pronounced reduction across the sample, suggesting that the recovered magnetization is largely aligned out-of-plane and oriented opposite to the external field.
We also perform a two-dimensional confocal scan of the stray field near the sample (Fig.~\ref{fig:fig4}c), where a large portion of the sample area exhibits a negative field. 
Fig.~\ref{fig:fig4}c illustrates the evolution of the maximum stray magnetic field as a function of pressure during both pressurization and depressurization, where a pressure-driven magnetic transition at around $0.5~$GPa is apparent. 
This transition in Cr$_{1+\delta}$Te$_2$ can be explained by pressure-induced reduction in atomic spacing that alters the overlap of orbitals and weakens the ferromagnetic exchange interaction strength, agreeing with density functional theory predictions (see Methods)~\cite{lin2018pressure,sun2018effects}.

\emph{Outlook} --- Looking forward, our work opens the door to a few intriguing directions.
On the application front, the close proximity of our 2D sensors and their demonstrated ability to simultaneously image stress and magnetism under high-pressure conditions enable the exploration of a wide range of pressure-induced phenomena. 
These include superconductivity and magnetism in materials science, as well as mechanical deformation and paleomagnetism in geology.
For instance, one can integrate \vbm sensors with the recently discovered thin-film nickelate superconductors to spatially correlate filamentary superconductivity (probed through Meissner repulsion) and local stress environment with sub-micrometer resolution.
On the scientific front, an open question remains as to how the pressure limit of \vbm sensors can be further extended. 
Insights from NV centers suggest that replacing NaCl with more hydrostatic pressure-transmitting medium, such as Daphne oil or noble gases, could substantially enhance the pressure operation limit.
Additionally, from a theoretical perspective, modeling the inter-system crossing and spin properties of \vbm centers under various stress configurations promises to yield valuable insights.

\setcounter{figure}{0}   
\renewcommand{\thefigure}{E\arabic{figure}}

\begin{center}
    \Large
    \textbf{Methods}
\end{center}
\section{DAC sample preparation}
In our experiment, we use a miniature Pasternak DAC~\cite{sterer1990multipurpose} to apply pressure. The cell body and backing plates are made of beryllium copper (BeCu). Between two backing plates are two type IIa diamond anvils with a $400~\mu$m culet diameter, compressing a stainless steel gasket. The gasket is preindented to $18~$GPa and laser drilled a hole with a diameter of $133~\mu$m. The sample chamber is insulated with a mixture of boron nitride (BN) powder and epoxy. Sodium chloride (NaCl) pressure medium is loaded to provide a quasi-hydrostatic environment. The hBN flake and Cr$_{1+\delta}$Te$_2$ magnet is picked up using polydimethylsiloxane (PDMS), and transfered then stacked onto a diamond anvil. A platinum wire of $50~\mu$m in width is placed across the heterostrucure are to deliver microwave drive for ODMR measurements.

\section{hBN sample preparation}
To grow the h$^{10}\text{B}^{15}\text{N}$ crystals, a mixture of 48.1 wt \% chromium (99.99\%), 48.1 wt \% nickel (99.9\%), and 3.8 wt \% boron-10 enriched (98\% B-10, 99.99\%) pieces were loaded into an alumina crucible (98\%) and covered before being loaded into a horizontal alumina (98\%) furnace. The tube was vacuumed to $< 80$~microtorr using a rotary vein pump. Then the tube was purged twice with N$_2$ and argon (99.9999\%) to 500~torr, then evcuated back to $< 80~$microtorr. Finally, the tube was filled to 90\% (approximately 780~torr) $^{15}$N$_2$ (98\% enriched, 99.9999\%) and complimented to 850~torr with H$_2$ (99.9999\%) before being closed up- and downstream of the tube.
The furnace was brought to 1550~C at 200~C/h then left to dwell for 24~hours at that temperature. This step was followed with a slow cool at 1~C/h to 1500~C. The furnace was then cooled at 50~C/h to 1350~C followed by 100~C/h to room temperature.

The sample boule was unloaded and removed from the alumina crucible. Heat release tape was used to exfoliate the crystals from the surface of the boule. After heating the tape to 135~C, acetone (99\%) was used to remove the free-standing crystals. The resulting free standing crystals were mounted between 2 1-inch square quartz plates and pressed under aluminum foil. These samples were irradiated at Ohio State University under fluences of $1.4\times10^{16}$ (1-hour) neutrons per square centimeter.

\section{C{r}$_{1+\delta}$T{e}$_2$ synthesis and characterization}
Chromium telluride Cr$_{1+\delta}$Te$_2$ nanoplates/nanoflakes with $\delta \approx 0.5$ were grown using a chemical vapor deposition method in a three-zone quartz tube furnace~\cite{coughlin2020near,coughlin2021van,coughlin2023extreme}. CrCl$_3$ (Alfa Aesar, purity of 99.9\%) and Te (Alfa Aesar, purity of 99.999\%) powders were used as precursors, each placed in a separate alumina boat within a smaller-diameter quartz tube. The Te boat and CrCl$_3$ boat were positioned in the center of zone 1 and zone 2, respectively. Several SiO$_2$ (300~nm)/Si substrates were placed downstream from zone 2 to zone 3. The temperatures of zones 1, 2, and 3 were set to be at 500~C, 750~C, and 750~C, respectively. A carrier gas mixture consisting of 1\% H$_2$ and 99\% Ar was supplied at a flow rate of 25~sccm, and the pressure was maintained at 50~Torr during the heat up process. As soon as the target temperatures were reached, the system was pumped down and the furnace was turned off and cooled down to room temperature gradually. 

\figTEM
TEM studies were conducted using 120~kV JEOL JEM 1400plus. Fig.~\ref{fig:TEM}a is a low magnification TEM image of a chromium telluride nanoplate. Selected area electron diffraction (SAED) patterns were collected along [001] and [112] zone axes, which are shown in Fig.~\ref{fig:TEM}b and c, respectively. The main diffraction spots in both patterns can be indexed by the structure of Cr$_{1+\delta}$Te$_2$ with $\delta \approx 0.5$.

\figAFM
The thickness of the Cr$_{1+\delta}$Te$_2$ sample is determined using atomic force microscopy (AFM). Fig.~\ref{fig:AFM} presents a 2D deflection scan obtained via AFM, along with a linecut measurement across the sample. Although the exact thickness of the flake studied in the DAC was not directly measured, the thickness of several other similar flakes from the same synthesis batch was calibrated. These measurements reveal a thickness range of 50–100~nm.

\section{First-principles calculations}
Density functional theory (DFT) calculations are carried out using the projected augmented wave method~\cite{blochl1994projector,kresse1999ultrasoft} as implemented in the Vienna \emph{Ab initio} Simulation Package. 
The Perdew-Burke-Ernzerhof~\cite{perdew1996generalized} of the generalized gradient approximation is adopted for the exchange-correlation functional. 
Here, we focus on the calculation of CrTe$_2$ instead of Cr$_{1+\delta}$Te$_2$ to reduce computational complexity.
An energy cutoff of 500~eV was set, and 18 $\times$ 18 $\times$ 8 k-point mesh was used to sample the entire Brillouin zone. 
Structural optimization is performed until the forces on all atoms are below 0.01~eV/$\mathrm{\AA}$, and the energy criterion for convergence in the self-consistent field calculation is set to $10^{-6}~$eV. 
The DFT+$U$ approach~\cite{anisimov1997first,dudarev1998electron} is employed to account for the on-site Coulomb interaction in the Cr $3d$ orbitals, with $U$ values of 5~eV and 6~eV being considered.
The exchange constant $J$ is calculated as a function of applied pressure. Specifically, $J$ is determined by calculating the energy difference between the ferromagnetic (Fig.~\ref{fig:DFT}a) and antiferromagnetic (Fig.~\ref{fig:DFT}b) configurations using a 2 $\times$ 2 $\times$ 1 extended structure, following the method described in~\cite{liu2022structural}. The calculated $J$ values, as shown in Fig.~\ref{fig:DFT}c, show a trend where increasing pressure results in a reduction in the exchange constant.
In the mean field approximation, the Curie temperature $T_C$ is proportional to the exchange coupling constant $J$. Therefore, the pressure dependence of $J$ in Fig.~\ref{fig:DFT}c suggests that the $T_C$ of the system is suppressed by increasing pressure.

\figDFT

\vspace{1mm}

\emph{Note added}: During the completion of this work, we became aware of a complementary work exploring \vbm in DAC to image pressure-induced magnetism of CrTe$_2$ in heterogeneous device (Vincent Jacques, private communication), which will appear in the same arXiv posting \cite{mu2025magnetic}.

\vspace{1mm}

\emph{Acknowledgments}: We gratefully acknowledge Vincent~Jacques, Zhao~Mu, Tristan~Club-Provost, Wonjae~Lee, Khanh~Pham, and Benchen~Huang for helpful discussions. We thank Shiling~Du for their assistance in experiment. This work is supported by NSF NRT LinQ 2152221, NSF ExpandQISE 2328837, and the Center for Quantum Leaps at Washington University. Support of T.~Poirier and J.~H.~Edgar for hBN crystal growth is provided by the NSF under Grant No. 2413808 and the Office of Naval Research under award N00014-22-1-2582. The work performed at Indiana University (chromium telluride growth, TEM characterization and DFT calculations) was supported by the NSF DMR-2327826.

\vspace{1mm}

\emph{Author contributions}--- C.Z. conceived the idea. G.H., R.G., Z.W., Z.L., A.L.R., C.Y. and B.Y. performed the high-pressure hBN measurement and analyzed the data. G.H., Z.W., Z.R. and S.R. prepared the high-pressure DAC. M.C. and X.W. helped on the heterostructure assembly. T.P. and and J.H.E. grew the hBN samples. J.H., T.Z. and S.Z. synthesized and characterized the chromium telluride nanoflakes and performed DFT calculations. N.Y.Y. and C.Z. supervised the project. G.H., R.G. and C.Z. wrote the manuscript with inputs from all authors.

\bibliographystyle{naturemag}
\bibliography{ref.bib}

\end{document}